# A Performance Evaluation of Container Technologies on Internet of Things Devices

Roberto Morabito
Ericsson Research, NomadicLab, Jorvas, Finland
roberto.morabito@ericsson.com

*Abstract*— The use of virtualization technologies in different contexts – such as Cloud Environments, Internet of Things (IoT), Software Defined Networking (SDN) – has rapidly increased during the last years. Among these technologies, container-based solutions own characteristics for deploying distributed and lightweight applications. This paper presents a performance evaluation of container technologies on constrained devices, in this case, on Raspberry Pi. The study shows that, overall, the overhead added by containers is negligible.

*Keywords*— *Internet of Things; Virtualization; Container; Docker; Performance; Benchmark; Power Consumption;*

## I. INTRODUCTION

Container-based virtualization can be considered a lightweight alternative to hypervisor-based virtualization. Containers implement isolation of processes at the operating system level of the host machine, thus, avoiding the overhead due to virtualized hardware and virtual device drivers. A container can be considered a tiny and isolated virtual environment, which includes a set of specific dependencies needed to run a specific application. The concept of "containerization" is not new in the virtualization world, but it has achieved more relevance and real-world adoption recently with the advent of Docker[1]. Docker introduces an underlying container engine, together with a functional API that allows easily building, managing, and removing a containerized application. Because of the small overhead produced, multiple containers can run even in devices with limited computation resources such as *Single Board Computer* platforms. These lightweight and versatile characteristics have facilitated the use of containers in different contexts ranging from *Cloud Computing* to *Internet of Things* (IoT) scenarios. Practical examples of the use of Docker containers in constrained devices can be found in [1] and [2]. In a *Capillary Network* [1], containers are executed in constrained environments without hardware virtualization support (e.g. a capillary gateway), and used for packaging, deployment, and execution of software suitable for network management and data pre-processing (filtering, compression and aggregation). In [2], the authors introduce a scale model of a Data Center composed of clusters of 56 Raspberry Pi devices, emulating every layer of a cloud stack from resource virtualization (implemented by means of Linux containers) to network components.

Considering the potential benefits introduced by containers, together with the significant increase of use cases, our study aims to assess the use of Docker containers in constrained environments, providing a detailed performance analysis.

## II. METHODOLOGY

The main goal of our empirical investigation is to evaluate, by the means of different benchmark tools, the performance of Docker when running on a *Single Board Computer* device such as *Raspberry Pi 2*[1] (RPi2). The *native* performance, i.e. running the benchmark tool without including any virtualization layer, is used as a reference for comparison. We used the Docker version 1.8.0 and, as operating system for the RPi2, the image provided by Hypriot[3] running *Raspbian Jessie* with Linux 3.18.11. The results are averaged over 15 runs. For the Docker results, table I shows the difference compared to the native test. The power consumption of the RPi2 during the execution of each single test is also reported. This is measured by using an external power meter, and a setup similar to the one used in [3]. Figure 1 shows the entire testbed environment setup.

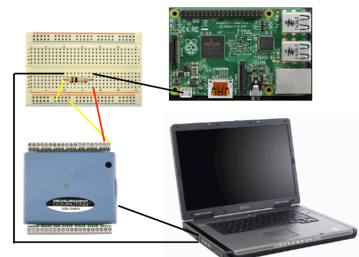

Fig. 1. Testbed setup.

## III. MEASUREMENT RESULTS

### A. Synthetic Benchmark

Synthetic benchmarks allow generating different types of workloads to challenge a specific sub-system of the hardware under test. *CPU*, *Memory I/O*, *Disk I/O,* and *Network I/O* are the main components that we want to test in this part of the experiments. The *sysbench*[4] tool allows performing multi-threaded tests for evaluating different parameters under intensive load. We used sysbench to test CPU, Disk I/O, and On-Line Transaction Processing (MySQL) performance.
**CPU.** The results prove an existing difference between the native case and the Docker case. However, the container engine introduces a negligible impact on the CPU performance, with a percentage difference in the order of 2.67%.
**Memory I/O.** To test the memory performance, we use the Unix command *mbw*[5]*,* which determines the available memory bandwidth by copying large arrays of data in memory, and performing three different tests (*memcpy*, *dumb*, *mcblock*). Similarly to the CPU case, native and container performance can be considered comparable, with a max percentage difference of 6,04% during the *memcpy* test.
**Disk I/O.** To evaluate Disk I/O performance we again use sysbench. The benchmark is set in order to execute random read/write operation. The outcome of this evaluation shows a performance degradation of Docker compared to the native case. This difference remains in the order of roughly 10%.

This work is partially funded by the FP7 Marie Curie Initial Training Network (ITN) METRICS project (grant agreement No. 607728). The author would like to thank the Cloud team of NomadicLab for providing help and support.

[1] http://www.docker.io
[2] http://www.raspberrypi.org/products/raspberry-pi-2-model-b/
[3] http://blog.hypriot.com/
[4] http://manpages.ubuntu.com/manpages/wily/en/man1/sysbench.1.html
[5] http://manpages.ubuntu.com/manpages/wily/en/man1/mbw.1.html

TABLE I. BENCHMARK RESULTS

| | CPU Benchmarking | | Memory Benchmarking | | | Power Consumption (W) |
|---|---|---|---|---|---|---|
| | Execution Time (seconds) | Power Consumption (W) | Average Speed (MiB/s) | | | |
| | | | memcpy | dumb | mcblock | |
| Native | 434.074 | 1.4140 | 598.22 | 70.93 | 601.55 | 2.2314 |
| Docker | 446 (+2.67%) | 1.4054 | 562.05 (+6.04%) | 70.43 (+0.7%) | 570.51 (+5.15%) | 2.2478 |
| | Network I/O Benchmarking | | Disk I/O Benchmarking | | | Power Consumption (W) |
| | Power Consumption (W) | | Operations Performed (MB) | | | |
| | TCP Client | TCP Server | Read | | Write | |
| Native | 2.2073 | 2.1597 | 123.25 | | 82.172 | 1.4328 |
| Docker | 2.2237 (+0.74%) | 2.2657 (+4.90%) | 107.062 (+13.13%) | | 74.719 (+9.07%) | 1.4321 |
| | UDP Client | | UDP Server | | Apache 2 Benchmarking (200 clients) – Power Consumption (W) | |
| | 100 Mbps | 80 Mbps | 100 Mbps | 80 Mbps | 5000 requests | 25000 requests | 100000 requests |
| Native | 2.1380 | 2.1358 | 2.1532 | 2.1310 | 2.4681 | 2.4719 | 2.4893 |
| Docker | 2.2974 (+7.4%) | 2.146 (+0.47%) | 2.210 (+2.65%) | 2.142 (+0.51%) | 2.3588 (-4.63%) | 2.3804 (-3.84%) | 2.3642 (-5.29%) |

**Network I/O.** *Iperf*[6] is a tool with predefined tests to measure network performance between hosts, generating bidirectional data transfer of both TCP and UDP traffic. In our investigation, we performed bidirectional tests, and the *iperf server* and *iperf client* are running in the RPi2 inside Docker containers. The scope of the network performance analysis is to quantify the power consumption of the RPi2, while performing intensive network workload when the same amount of traffic is exchanged between the RPi2 and another host. The decision to execute the test in both directions is due to the fact that TCP and UDP have different code paths for sending and receiving traffic. In the case of TCP traffic, when the RPi2 acts as client, we do not observe any power consumption mismatch between the two cases. A slight increase of power consumption (+4.90%) can be observed when the container is receiving TCP traffic. With UDP traffic, we repeat the test by fixing two different throughput values. This is due to the fact that for 100 Mbps (link completely saturated), a not negligible amount of packets loss is observed. Reducing the max throughput to 80 Mbps, there are not packets loss. In this latter case, performance by using containers are aligned to the native case both in the client and in the server case; this results show a difference in respect to the TCP analysis.

### B. Application Benchmark

The aim of the application benchmark experiments is to emulate potential real-world applications.

**MySQL.** This test was executed to benchmark real database performance. We created two databases with one and two million rows, respectively, and varied the number of concurrent threads. Figure 2 depicts the results of the performed number of transactions per second with the increasing number of threads. The corresponding power consumption is also reported. From the graph can be observed how there is no a relevant difference in terms of performance between the native case and the Docker configuration, especially for a high number of concurrent threads. This represents a relevant result because of the characteristics of the specific workload.

**Apache2.** We emulated the behavior of a busy server that serves an increasing number of concurrent clients (100, 200, and 300). The Apache server – which is running in the RPi2 – has to handle an increasing number of requests executed by the clients during the benchmarking session (*Apache HTTP server benchmarking tool*[7] was used). The power consumption results show an interesting aspect. Docker produces lower power consumption compared to the native case. Investigating more on the reasons of such result, we analyzed two relevant metrics: the number of request per second handled by the server and the execution time. In Figure 3, we can observe how the native execution outperforms Docker in any comparable case. Hence, it is interesting to observe this trade-off between power consumption and performance. The *Apache server* containerization implies lower power consumption at the expense of a slight degradation in performance.

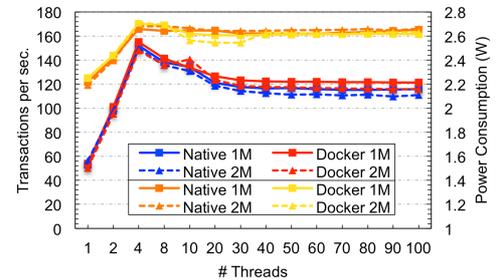

Fig. 2. MySQL Transactions per second (red, blue) and power consumption (yellow, orange) vs. concurrency.

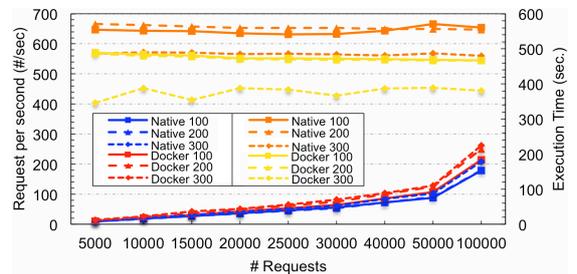

Fig. 3. Apache2 Request per second (yellow, orange) and execution time (red, blue) vs. number of requests

### IV. CONCLUSIONS AND FUTURE WORK

This paper introduces a performance evaluation of container-based technologies running on top of a Raspberry Pi 2. Results show an almost negligible impact of the container virtualization layer in terms of performance, if compared to native execution. As future work, we will conduct similar experiments with additional *Single Board Computer* platforms.

[6] http://iperf.sourceforge.net/
[7] https://httpd.apache.org/docs/2.2/programs/ab.html